Pressure-driven flow of a micro-polar fluid: measurement of the velocity profile.

Peters F., Lobry L. and Lemaire E.
CNRS - Université de Nice
LPMC - UMR 6622
06108 Nice cedex 2 - France

**Synopsis**

The pressure-driven flow of a suspension of spinning particles in a rectangular channel is studied using an acoustic method. The suspension is made of insulating particles (PMMA) dispersed in a slightly conducting oil ( Ugilec + Dielec) and is subjected to a DC electric field. In such a case, the particles are polarized in the direction opposite to that of the electric field and begin to rotate in order to flip their dipoles in the field direction. Such a rotation of the particles is known as Quincke rotation and is responsible for an important decrease of the effective viscosity of the suspension. Indeed, due to the electric torque exerted on the particles, the stress tensor in the suspension is not symmetric anymore and a driving effect arises from the anti-symmetric part. When such a suspension flows through a rectangular channel, the velocity profile is expected to deviate from the usual Poiseuille flow.

In this paper, the velocity profiles are measured using Pulsed Ultrasound Doppler Velocimetry technique. They compare well with those that are computed from the otherwise measured rheological law.

**I. INTRODUCTION**

In this paper, we are interested in the rheological behaviour of a suspension of spheres which experience a torque. In such a case, it is well known that the stress tensor is no longer symmetric but that an anti-symetric part has to be added to the usual symmetric stress tensor. The role played by this anti-symmetric stress on the mechanical behaviour of suspensions has been widely studied both theoretically and experimentally. Such suspensions are known as micro-polar fluids and their theory developed by Eringen (1966) and Dahler, Condiff and Scriven [Dalher and Sriven (1961), Condiff and Dalher (1964)] around 1960. The mechanical behaviour of such fluids, is given by two vector fields, the linear velocity field as in the case of non polar fluids and the particle angular velocity field [Lukhaszewicz (1999)]. These fields are the solutions of two coupled equations: the linear momentum and internal angular-momentum balance equations. In recent years, the theory of micropolar fluids has received a new interest with the development of microfluidics and ferrofluid rheology [Rosensweig (2007)]. The ferrofluids submitted to alternating or rotating magnetic fields are undoubtedly the most famous micro-polar suspensions. The external magnetic field exerts a torque on the dipole moments carried by the particles which consequently rotate with an angular velocity different from that of the surrounding liquid. Many theoretical [Perez-Madrid et al. (1999); Rinaldi et al. (2002); Korlie et al. (2008); Shliomis and Konstantin (1994)] and experimental [Bacri et al. (1995); Rinaldi et al. (2005)] investigations have addressed the role of the magnetic torque on the rheological behaviour of ferrofluids. In particular, it has been shown that the application of a fast enough oscillating magnetic field forces the particles to rotate faster than the host fluid leading to a reduction of the apparent viscosity of the ferrofluid [Bacri et al. (1995)]. A true apparent "negative" viscosity can be obtained upon submitting the ferrofluid to a rotating magnetic field [Rinaldi et al. (2005)].

In this paper, we propose another way to induce a torque on the suspended particles. It consists in applying a DC electric field to a suspension of insulating particles dispersed in a slightly conducting liquid. Indeed, when a non conducting particle immersed in a semi-insulating liquid is submitted to a sufficiently high amplitude DC field, it begins to rotate spontaneously around itself. This instability is known as Quincke rotation [Quincke (1896); Jones (1995)]. When the suspension is motionless and a DC field is applied, the particles start rotating around themselves in any direction perpendicular to the field and the average spin rate of the suspension is zero. But, when a velocity gradient is applied along the E field direction, the particles rotation axis is favoured in the vorticity direction. Therefore, the degeneracy of the rotation direction is eliminated, giving rise to a non zero spin rate of the ensemble of particles. This macroscopic spin rate may drive the suspending liquid and thus lead to a decrease of the apparent viscosity of the suspension [Lemaire et al. (2008)].

We have been working for the last years both on the modelling and on the experimental characterisation of the influence of Quincke rotation on the rheological behaviour of suspensions. In particular, we have been able to propose a constitutive relation for the rheological behaviour of such a suspension of spinning particles. The model predicts that, for low particle concentrations ($\phi<1/6$), there exists a bistable flow regime and the apparent viscosity of the suspension is negative when it is submitted to a small shear rate. Experimentally, we were not able to show evidence of this negative apparent viscosity but we measured a huge viscosity decrease [Lemaire et al. (2008)]. Moreover, we have shown that when an electric field was applied to such a suspension flowing in a rectangular channel, the flow rate could be enhanced. The flow rate increase is expected to be accompanied by a change in the velocity profile shape since, when an electric field is applied, the rheological behaviour of the suspension is not Newtonian anymore. The aim of the present paper is to present the direct measurement of the velocity profiles and their modification induced by the electric field. In section II, we present briefly the mechanism involved in the viscosity reduction. The third part is devoted to the description of the experimental set-up and method. The results are presented in section IV and discussed in section V.

## II. QUINCKE ROTATION-INDUCED VISCOSITY REDUCTION

All the details concerning the possibility of decreasing the viscosity of a suspension thanks to DC-electro-rotation have been previously given [Lemaire et al. (2008)]. Here we just mention the starting point of the analysis and the main results.

### A. Quincke rotation of a single particle

The electroration of a particle induced by a DC electric field is known as Quincke rotation and occurs when an insulating particle immersed into a low conducting liquid is subjected to a sufficiently high DC field. In such a situation, as first observed by Quincke at the end of the nineteenth century, the particle can rotate spontaneously around itself along any axis perpendicular to the electric field. A simple physical interpretation has been given by Secker and Scialom (1968) and refined by Melcher (1974): during their migration, the free negative and positive charges of the liquid meet the insulating particle and accumulate at its surface. This results in a charge distribution with the particle polarised in a direction opposite to the field. This equilibrium is unstable and when the particle begins to rotate in order to flip the orientation of the induced dipole, it undergoes a driving torque. As stated by Jones (1984), the Quincke rotation may take place only if the charge relaxation time, which is given by the ratio of the dielectric constant to the conductivity, is lower in the liquid than in the particle. In such a case, the retarded part of the dipole, **P**, which is associated with the charge distribution at the particle/liquid interface, is in the opposite direction from the electric field. Then, if the particle is slightly rotated, the deviation of its dipole moment **P** produces a torque, $\mathbf{\Gamma^E}=\mathbf{P}\times\mathbf{E_0}$, which tends to increase the angular tilt further.

It has been shown [Cebers (1978); Pannacci et al. (2007a)] that the retarded part of the dipole moment was given by a relaxation equation:

$$\frac{\partial \mathbf{P}}{\partial t} = (\boldsymbol{\omega}\times\mathbf{P}) - \frac{1}{\tau_M}\left(\mathbf{P}-(\chi^0-\chi^\infty)\mathbf{E_0}\right) \qquad (1)$$

Where $\boldsymbol{\omega}$ is the particle angular velocity and $\tau_M$, the Maxwell time that is the characteristic relaxation time of the dipole moment:

$$\tau_M = \frac{\varepsilon_p + 2\varepsilon_l}{\gamma_p + 2\gamma_l} \qquad (2)$$

$\gamma_i$ and $\varepsilon_i$ being the electric conductivity and the dielectric permittivity of the liquid (i=l) and of the particle (i=p).

$\chi^0$ and $\chi^\infty$ are the polarisability factors of the particle (radius a) at low and high frequency respectively:

$$\chi^0 = 4\pi\varepsilon_l a^3 \frac{\gamma_p-\gamma_l}{\gamma_p+2\gamma_l} \quad \text{and} \quad \chi^\infty = 4\pi\varepsilon_l a^3 \frac{\varepsilon_p-\varepsilon_l}{\varepsilon_p+2\varepsilon_l}$$

Besides the electric torque, $\mathbf{\Gamma^E}=\mathbf{P}\times\mathbf{E_0}$, the particle is subjected to a viscous torque:

$$\mathbf{\Gamma^\eta} = -\alpha(\boldsymbol{\omega}-\boldsymbol{\omega_0}) \qquad (3)$$

where $\alpha=8\pi\eta_l a^3$ is the rotational friction coefficient of the spherical particle, $\eta_l$ being the dynamic viscosity of the liquid and $\boldsymbol{\omega_0}$, the angular velocity of the surrounding fluid (in a quiescent fluid, $\boldsymbol{\omega_0}=\mathbf{0}$).

The particle spin rate is obtained upon solving equation (1), together with the angular momentum principle:

$$I\frac{d\boldsymbol{\omega}}{dt} = \mathbf{P}\times\mathbf{E_0} - \alpha(\boldsymbol{\omega}-\boldsymbol{\omega_0}) \qquad (4)$$

where I is the particle moment of inertia.

When $\omega_0=0$, the Quincke rotation takes place only if the electric field intensity is high enough for the electric torque to overcome the resistant viscous torque. The threshold intensity is expressed as:

$$E_C = \sqrt{\frac{-\alpha}{\tau_M(\chi^0-\chi^\infty)}} \quad (5)$$

**B. Quincke rotation in a suspension**

If we are interested in the rotation of an ensemble of particles in a suspension subjected to a constant shear rate $\dot{\gamma}$ (figure 1), using a mean field approach, $\omega_0$ is taken as the average angular velocity of the suspending liquid [Brenner (1970)]:

$$-\dot{\gamma}\mathbf{e_x} = 2\left(\phi\boldsymbol{\omega}+(1-\phi)\boldsymbol{\omega_0}\right) \quad (6)$$

where $\phi$ is the particle volume fraction.

When the applied electric field intensity is higher than a critical value, $E^* = \frac{E_c}{\sqrt{1-\phi}}$, the coupled equations (4) and (6) admit several stationary solutions [Lemaire et al. (2008)]. In the last reference, thanks to a linear stability analysis, we have been able to propose a criterion for the stability of the solutions: if the particles and the surrounding liquid rotate in opposite direction, the solution is unstable.

Once the particle angular velocity is known, it is injected in the model proposed by Howard Brenner (1970) to relate the effective viscosity of a suspension to the particle spin rate:

$$\eta_{\text{eff}}(\phi) = \eta(\phi) + \eta_l \frac{\frac{3}{2}\phi}{1-\phi}\frac{\frac{\dot{\gamma}}{2}+\omega_x}{\frac{\dot{\gamma}}{2}} \quad \text{with} \quad \eta(\phi)=\eta_l\frac{1+\frac{3}{2}\phi}{1-\phi} \quad (7)$$

where $\eta_l$ is the fluid viscosity. From this expression, it is clear that when the particles rotate faster than the liquid ($|\omega_x| > \dot{\gamma}/2$) due to the electric torque, the effective viscosity of the suspension is lowered. It should be noted that the second term of the right hand side of eq.(7) is related to the anti-symmetric part of the stress tensor which is responsible for the driving effect.

Figure 2.a shows the expected variation of the normalised viscosity, $\eta_{eff}(\phi)/\eta(\phi)$, with the reduced shear rate, $\dot{\gamma}\tau_M$, for a suspension whose solid volume fraction is 5%. It should be stressed that when the shear rate is lower than a critical value, $\dot{\gamma}^*$, the effective viscosity, $\eta_{eff}(\phi)$, is expected to be negative [Lemaire (2008)].

Here we focus our attention on the flow of such a suspension of spinning particles trough a rectangular channel. In such a flow, the control parameter is not the shear rate but the shear stress which is null at the centre of the channel and maximum at the walls. The relation between the shear rate and the shear stress is shown on figure 2.b. The regions of the rheogram that are explored in a capillary flow are those where the absolute value of the shear rate is larger than $\dot{\gamma}^*$.

To calculate the velocity profile in the rectangular capillary, we suppose that the capillary width (along x-direction), $\ell$, is much larger than its thickness (along z-direction), e, so that the velocity and the stress tensor, except for the pressure terms, depend only on the variable z (see figure 3). In that case, the pressure, p, depends only on y.

The balance equation for linear momentum reads:

$$-\frac{dp}{dy} + \frac{d\sigma}{dz} = 0 \qquad (8)$$

where $\sigma_{yz}(z) = \sigma(z)$ is the shear stress

Equation (8) simplifies by supposing that $\sigma(-z) = -\sigma(z)$:

$$\sigma = \frac{dp}{dy} z \tag{9}$$

Once the rheological constitutive law which depends on the electric filed intensity: $\dot{\gamma} = f_E(\sigma)$, is deduced from equation 6, the velocity profile can be computed:

$$v_y(z) = \int_{-\frac{e}{2}}^{z} \dot{\gamma}(z_1)\, dz_1 = \int_{-\frac{e}{2}}^{z} f_E(\frac{dp}{dy} z_1) dz_1 \tag{10}$$

Figure 4 shows the profiles deduced from the expression (7). As expected Quincke rotation is responsible for an increase of the suspension linear velocity and, as the field intensity increases, the velocity profile becomes sharper. The singular point at z=0 originates in the cancellation of the shear stress (or equivalently of the effective viscosity) for a finite shear rate. Indeed, at z=0, the shear stress equals zero (see equation (9)) while the shear rate is expected to be $+\dot{\gamma}^*$ at z=0$^+$ and $-\dot{\gamma}^*$ at z=0$^-$. This discontinuity in the shear rate is accompanied by an inversion of the particle angular velocity (see figure 3 and equation (7)).

In the following we present some experiments which have been carried out in order to verify these predictions.

**III. EXPERIMENTAL SET-UP**

The velocity profile of a suspension flowing through a rectangular capillary is measured thanks to a method of ultrasonic speckle velocimetry. The experimental device is presented on figure 5.

**A. Materials**

The suspensions is made of PMMA particles (Microbeads CA6-6) (diameter 6 µm, $\varepsilon_2 = 2.4\varepsilon_0$, density $\rho_p = 1.18\ 10^3 \text{kg.m}^{-3}$) dispersed in a mixture of transformer oil (Dielec S, Hafa France) and Ugilec (Elf Atochem). The following properties of the suspending liquid have been measured at T=22°C: $\varepsilon_l=3.7\varepsilon_0$, $\eta_l=16.5$ mPa.s. Its density $\rho_l=1.14\ 10^3$kg.m$^{-3}$ is close to that of the particles. Its conductivity has been raised to $\gamma_l=5.4\ 10^{-8}$ S.m$^{-1}$ upon addition of an ionic-surfactant (AOT salt, sodium dioctylsulfosuccinate, Sigma Aldrich). The particle conductivity is small enough ($\gamma_1 \approx 10^{-14}$ S.m$^{-1}$) to consider them as insulating. Accounting for the electric characteristics of the particles and of the suspending liquid, we deduce the values of the critical field and of the Maxwell time: $E_c$=1800V/mm and $\tau_M$=1.6ms. The volume fraction of the solid particles is $\phi=5.10^{-2}$ and the viscosity of the suspension, $\eta_s$=19.3 mPa.s., has been measured with a controlled stress rheometer Carrimed CSL 100. The particle volume fraction has been chosen in such a way that the effect of Quincke rotation on the material rheology is sufficiently high, without inducing too much ultrasound multiple scattering that would have made the measurement of the velocity profile impossible.

**B. Flow geometry**

The suspension flows in a rectangular channel, $\ell$ =1.5 cm in width and e=1 mm in thickness. The upper and lower walls of the tube are made of PET films, 175 microns in thickness, coated with a conducting ITO layer and glued on a rigid plexiglass plate. The coated PET films serve also as electrodes so that the electric field is applied along the velocity gradient direction. They are connected to a high voltage supply (Trek 610E) that is fed by a 0.5 Hz square wave signal from a generator (Agilent 33120A). Two holes in one electrode wall, L=22 cm apart, 2mm in diameter, let the suspension flow in and out of the tube. The pressure drop across the channel is controlled by placing the tank containing the suspension at a height h=17.5cm above the capillary which lies on an horizontal plane.

**C. Pulsed Ultrasonic Doppler Velocimetry**

The velocity profile measurement method is very similar to that precisely described in [Manneville et al. (2004)]. The principle of the method is quite simple: a focused ultrasonic transducer emits a pulse wave toward the tube in which the suspension flows (figure 6.a). Each particle scatters the wave back to the transducer that works as a receiver as well. As shown in figure 6.b, if a particle moves with a velocity v and two pulses are sent at time interval T, their echoes will be received at a time interval T+$\tau_d$ where $\tau_d$ is related to the distance travelled by the particle during the time interval T:

$$\tau_d = \frac{2vT\sin\theta}{c} \qquad (11)$$

$\theta$ being the angle between the acoustic wave propagation direction and the normal of the capillary wall and c, the sound velocity in the suspending liquid.

As a consequence, measuring the time delay is equivalent to measure the particle velocity. Moreover, the z-position of the scattering particle in the tube is deduced from the lapse of time between the emitted and received pulses. The measurement of the delay time for all the positions in the tube gives the velocity profile in the medium.

Because a single pulse is scattered by a large number of particles, the received signal is the superposition of all scattered pulses, resulting in a speckle signal, each part of it corresponding to a position in the tube (figure 7). In order to process such waveforms and extract the velocity profile, two signals corresponding to two sent pulses are divided in windows of width $\Delta t$. If each window from a pulse is labelled by its central time k $\Delta t/2$ (two successive windows overlap), the $k^{th}$ window corresponds to a propagation distance in the suspension c k $\Delta t/2$, so that it can be regarded as the signal scattered by the particles that are at a distance ck$\Delta t/4$ from the tube wall. The delay time between the reception of two successive windowed pulses, $\tau_d$, is determined by the maximum of the correlation function of the windowed signals. Thus, only two successive speckle signals are in principle necessary to obtain the complete velocity profile. Actually, the velocity profile is averaged over 400 pairs of pulses.

The spatial resolution along the acoustic beam is given by the time window width $\Delta x = c\Delta t/4$. We chose $\Delta t = 100$ ns $\approx 3.6/f$ where $f = 36$ MHz is the central frequency of the emitted pulse, so that the spatial resolution is approximately the central wavelength of the pulse $\lambda \approx 40$ μm.

As shown on figure 5, a hole is made in the upper Plexiglas plate and the transducer is immersed in a water tank in order to avoid any large impedance mismatch. The broadband PVDF ultrasound transducer (Panametrics PI-50-2, actual central frequency 36MHz) has an active element of 6.35mm in diameter and a focal length of 12.7 mm, and the focal spot over which the beam has a constant section is approximately 1 mm long in the axial direction and 90 µm in lateral dimension. It is mounted on a precision rotation stage that allows the measurement of the angle $\theta_0$ between the transducer axis and the direction normal to the capillary wall. It can be translated thanks to a linear stage so that the focal spot coincides with the tube gap. The transducer is fed by a high voltage pulse from a pulser-receiver (Panametrics 5073 PR, ultrasonic bandwidth 75 MHz), which amplifies the back-scattered signal and sends it to a high frequency PCI digitizer board (Acqiris DP 235, 1GS/s, 0.5 GHz, 4Mb on-board memory). The digital signal is then loaded to the computer for further processing. The repetition frequency $1/T$ varies between 3kHz and 6 kHz, depending on the highest velocity in the tube.

The sound velocity in the suspending liquid is not known accurately, but it is not really necessary. First, the quantity that allows to extract the velocity from the delay time, $\tau_d$, is $c/\sin(\theta)$, that is related to the component of the wave vector parallel to the tube wall. It is well known that this quantity is continuous across any interface, so that $c/\sin\theta = c_0/\sin\theta_0$, where $c_0$ is the sound velocity in the water in which the transducer is immersed, and $\theta_0$ is the angle between the transducer axe and the tube wall on the water side. Both quantities are precisely known: $c_0=1488 m.s^{-1}$ at room temperature, $T=22°C$ and all the measurements were performed with $\theta_0=18°$.

The z-position of the particles in the capillary is deduced from the delay time between the emission and the reception of a single pulse. We assume that the distance between the upper capillary wall and the scattering particle is proportional to this delay time. As a consequence, only the times at which the pulse enters and exits the channel are to be known. In that purpose, the velocity profile versus delay time is fitted to a parabolic Poiseuille flow, when no electric field is applied, and the times at which the fitting curve cancels give the position of the walls.

## IV. EXPERIMENTAL PROCEDURE AND RESULTS

### A. Measurement of the profile in the absence of an electric field

During 30s, the suspension flows in the tube and is collected in a vessel for weighing. The acoustic measurement is performed after 15s. The level in the tank is controlled and h does not drop below 17 cm (h=17.5 cm at the beginning). The velocity profiles are first measured when no electric field is applied in order to determine the wall position and the hydraulic resistance of the suspension feeding system (tube from the reservoir to the capillary, connection of the tube with the capillary…).

First, the velocity profile measured without applying any electric field is fitted to a parabolic curve and the position of the wall is determined (figure 8). It should be stressed that the part of the curve close to the lower wall ($-e/2<z<0$) is affected by the sound reflection against the wall.

The viscosity of the suspension, $\eta_s=19.3 \cdot 10^{-3}$Pa.s has been measured with a controlled stress rheometer Carrimed CSL 100. Since for such a low particle volume fraction, $\phi=5\%$, the suspension exhibits a Newtonian behavior, the velocity profile in the channel is expected to be parabolic:

$$v_y = -\frac{e^2}{8\eta}\left(1-\left(\frac{2z}{e}\right)^2\right)\frac{dp}{dy} \tag{12}$$

leading to the flow rate:

$$Q = -\frac{\ell e^3}{12\eta}\frac{dp}{dy} = \frac{\ell e^3}{12\eta L}\Delta p \tag{13}$$

where L is the length of the tube, $\ell$ its width, e its thickness, and $\Delta p$ the pressure drop over the length of the capillary. The Reynolds number of the flow is taken as $R_e=\rho Q/(\eta \ell)\approx 2$. The entry length based on the thickness, e, of the pipe is $l_{ent}=R_e e \approx 4$mm. The entry length based on the width, $\ell$, of the channel can be computed as approximately 60 cm. So, the measurement of the velocity profiles 15 cm beyond the capillary entry is meaningful and the hypothesis of a 2D flow is quite correct.

The flow rate is computed from the velocity profile $Q_{ac}=5.8 \cdot 10^{-7}$m$^3$.s$^{-1}$ and compares well with the flow measured by weighing $Q=6.06 \cdot 10^{-7}$m$^3$.s$^{-1}$.

From $Q_{ac}$, the pressure gradient in the pipe can be evaluated and the hydraulic resistance of the rest of the circuit can be deduced:

$$R_h = \frac{\rho g h}{Q} - \frac{12\eta L}{\ell e^3} = 8.2\, 10^8 \text{Pa.m}^{-3}.\text{s} \tag{14}$$

R $_h$ is to be compared to the hydraulic resistance of the channel when no electric field is applied:

$$R_c = \frac{12\eta L}{\ell e^3} = 2.6\, 10^9 \text{Pa.m}^{-3}.\text{s} \tag{15}$$

Since $R_h$ is of the same order of magnitude as $R_c$, it is necessary to take it into account when the knowledge of the pressure drop across the capillary is needed. In particular, when an electric field is applied in the capillary, its hydraulic resistance is changed since the apparent viscosity of the suspension is reduced while $R_h$ is expected to be almost constant.

**B. Measurement of the profile in the presence of an electric field**

Rather than a true DC electric voltage, we apply a 0.5Hz square wave voltage to the electrodes, in order to avoid the polarization of the electrodes and the depletion of the electric charges in the tube. The time required to perform one measurement is smaller than one second. Thus, it is possible to measure the velocity profile for each polarity of the electric field. No general difference has been noticed between the two polarities, suggesting that no significant migration of the particles under the action of the electric field takes place. The mean profile is then computed. The velocity profiles for different electric field intensities are represented by the symbols on figure 9. As expected, the Quincke rotation of the particles results in a decrease of the effective viscosity, i.e. an increase of the flow rate, together with a deviation of the profile from the simple parabolic Poiseuille law.

**C. Comparison with rheometric data**

To check the consistency of this result with other rheometric experiments, we compute the expected velocity profiles from the relation between shear rate and shear stress that has been measured in a rheometer for a suspension submitted to the same E field intensities. The rheometric measurements were carried out with a controlled stress rheometer, Carrimed CSL 100, in Couette flow geometry. The coaxial cylinders also serve as electrodes so that the electric field is applied along the velocity gradient direction (radial direction). A high voltage is applied to the outer stator while the inner rotor is grounded to the earth through an electrode dipping in a mercury reservoir situated at the top of the rotation axis of the rheometer. The gap between the two cylinders (1 mm) is narrow compared to their radii (14 and 15 mm) so that the flow can be approximated by a simple shear flow. The variation of the viscosity with the shear rate is shown on figure 10 for the different intensities of the electric field. These experimental results agree qualitatively with the predictions of the model (figure 2.a). In particular, the concordance between experimental data and theoretical values of the viscosity is satisfactory at high shear rates. A discrepancy appears for low values of the shear rate: while the model predicts a negative effective viscosity, the measured viscosity remains positive and even exceeds its zero electric field value for the smallest shear rates. In a previous paper [Pannacci (2007b)] we have attributed this behaviour to a structuring of the suspension.

To compute the expected velocity profiles we need to know the pressure drop across the channel. Its value is obtained by writing that the total pressure drop is the sum of the pressure drop in the capillary and in the rest of the device:

$$\rho g h = L \frac{dp}{dy} + Q R_h \tag{16}$$

Then using equation (10), the flow rate is deduced:

$$Q = 2\ell \int_0^{\frac{e}{2}} \left( \int_{-\frac{e}{2}}^{z} f_E (\frac{dp}{dy} z_1) dz_1 \right) dz = \frac{\ell e^2}{2\sigma_w^2} \int_{\sigma_w}^{0} f_E(\sigma) \sigma \, d\sigma \tag{17}$$

where $\sigma_w = \frac{e}{2} \frac{dp}{dy}$ is the wall shear stress. Equations (16) and (17) together lead to the following equation:

$$\int_{\sigma_w}^{0} f_E(\sigma) \sigma \, d\sigma = \frac{2L}{R_h \ell e^2} \sigma_w^2 \left( \frac{2}{e} \sigma_w + \frac{\rho g h}{L} \right) \tag{18}$$

Using the rheometric data, eq.(18) is solved numerically to obtain the wall shear stress, $\sigma_w$. From $\sigma_w$, the pressure gradient is deduced and its value is injected in equation (10) to compute the expected velocity profiles. They are displayed together with the profiles obtained from acoustic measurement on figure 9 where a clear agreement should be noted.

## V. DISCUSSION

Thanks to ultrasonic speckle velocimetry, we have been able to measure the flow profiles of a suspension flowing in a rectangular channel and to evidence the role played by the particle electrorotation. The theoretical model predicts that the viscosity is zero when the suspension is subjected to a zero-shear stress, at the centre of the capillary. This leads to a singular point in the velocity profile. Experimentally, the velocity profiles do not exhibit any singularity but appear to be flatter than the parabola around $z = 0$. This is consistent with the rheometric measurements of figure 10 which, as noted in section IV. C, display a high viscosity value at very low shear rate. Nevertheless, the direct measurement of a negative effective viscosity or zero effective viscosity is difficult because it is supposed to arise when the shear stress is so low that it is hard to control it with a classical rheometer. Consequently, the measurement of the flow profile offers a promising way of measuring such a zero viscosity.

The measurement of the flow profile should also offer information on the possible effects of non-zero spin viscosity on the flow of such a micro-polar fluid. A non-zero spin viscosity introduces a coupling between the linear momentum and internal angular-momentum balance equations [Rosensweig (1997)]:

$$-\nabla p + 2\zeta(\phi)\nabla \times \boldsymbol{\omega} + (\eta(\phi) + \zeta(\phi))\nabla^2 \mathbf{v} = \mathbf{0}$$
$$\frac{\phi}{\frac{4}{3}\pi a^3}\mathbf{P} \times \mathbf{E}_0 + 2\zeta(\phi)(\nabla \times \mathbf{v} - 2\boldsymbol{\omega}) + \eta'(\phi)\nabla^2 \boldsymbol{\omega} = \mathbf{0} \quad (19)$$

where $\zeta$ is the vortex viscosity that actually appears in eq.(7) to account for the anti-symmetric part of the stress tensor:

$$\varsigma = \frac{\frac{3}{2}\phi}{1-\phi}\eta_l \qquad (20)$$

The spin viscosity η' is expected to be of the functional form: $\eta' = \eta_l a^2 f(\phi)$ [Rosensweig (1997); Feng et al. (2006)]. Consequently the internal angular momentum diffusion term in equation (19) depends on the parameter $(a/e)^2$. In our experiment, the value of this parameter ($(a/e)^2 \approx 10^{-5}$) is $10^6$ times larger than in the case of a ferrofluid flow ($(a/e)^2 \approx 10^{-11}$). Furthermore, according to the theoretical work of Rinaldi and Zahn (2002) or of Lukaszewicz (1999) where equation (19) is solved for various values of the ratio $\eta'/\zeta d^2$, the effect of the spin viscosity should be observable but weak in our experiment. An order of magnitude of this effect can be obtained by comparing the first and the third term of the second equation (19). From equations (1) and (5), the first term scales as $6\phi\eta_l \frac{\omega}{1+(\omega\tau)^2}\left(\frac{E}{E_c}\right)^2 \sim 6\phi\eta_l\omega$. The third term can be expressed as $\eta_l \left(\frac{a}{e}\right)^2 f(\phi)\omega$, so that the ratio of the spin diffusion term to the volume electric torque is given by: $\left(\frac{a}{e}\right)^2 \frac{f(\phi)}{6\phi}$. Following Rosensweig (Rosensweig,1997), $f(\phi) \sim 4\left(\frac{\pi}{6\phi}\right)^{2/3}$, leading to $10^{-3}$ for the sought ratio.

To catch such a small effect, the experiment has to be improved. The accuracy of the velocity measurement seems satisfactory although the use of a pulser-receiver with a higher signal to noise ratio would be valuable. We plan also to achieve a better control of the flow parameters. In particular, the true pressure gradient in the channel will be measured thanks a technique that we used previously and that consists in mounting pressure sensors on the capillary walls [Lemaire (2006)]. Moreover, in the experimental section (IV. A), the entry length based on the channel width ℓ, has been evaluated to be four times the channel length. ℓ should be increased to make this entry length much larger than the capillary length ensuring that the 2D approximation is correct at the channel centre. Finally, we plan to use a suspension with a suspending liquid whose conductivity is lower than in the present experiment. Then, the critical electric field intensity, $E_c$, which varies as the square root of the electrical conductivity (see section II.A) will be lowered and the intensities of the electric field we can apply without observing dielectric breakdown would be much higher than $E_c$ that is a condition for a significant change in the rheological behaviour of the suspension.

**ACKNOWLEDGEMENTS**

The authors acknowledge Professors Markus Zahn and Pierre Mills for fruitful discussions.

**Figure captions**

Figure 1: The suspension is subjected to both a DC electric field and a simple shear flow. All the particles rotate around themselves with an axis pointing in the vorticity direction. The particles rotate faster than the surrounding liquid, leading to a decrease of the apparent viscosity

Figure 2 Predicted rheological behaviour of a suspension containing a solid volume fraction of 5% for different electric field strengths. a) Normalised viscosity, $\eta_{eff}(\phi)/\eta(\phi)$, versus the reduced shear rate , $\dot{\gamma}\tau_M$. b) Normalised shear stress, $\sigma\tau_M/\eta(\phi)$, versus the reduced shear rate. It appears that for sufficiently high electric fields, the shear rate can have a finite value, $\dot{\gamma}^*$, while the applied shear stress is zero.

Figure 3: The flow of a suspension in a rectangular channel.

Figure 4: Theoretical velocity profiles of a suspension containing a solid volume fraction of 5% and flowing between two plates. The velocities have been normalised by the maximum velocity in absence of an electric field, $v_{max}(E=0)$.

Figure 5: Experimental device

Figure 6: Principle of Pulsed Ultrasonic Doppler Velocimetry measurement. a) Focalised ultrasonic beam.. b) Delay time, $\tau_d$, between two successive echoes.

Figure 7: Typical example of the scattered signal. .

Figure 8: Velocity profile obtained with a suspension containing a solid volume fraction of 5%. The symbols correspond to the measurements while the solid line is the parabolic fit.

Figure 9: Velocity profiles obtained for different electric field intensities. The symbols correspond to the acoustic measurements while the lines are the profiles deduced from the rheometric data.

Figure 10: Variation of the effective viscosity with the shear rate measured with a controlled stress rheometer in Couette geometry for different field intensities.

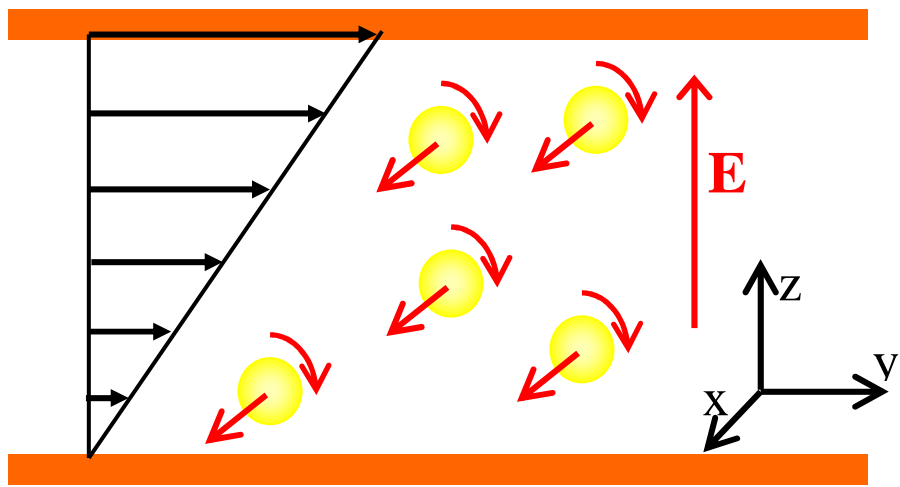

Figure 1

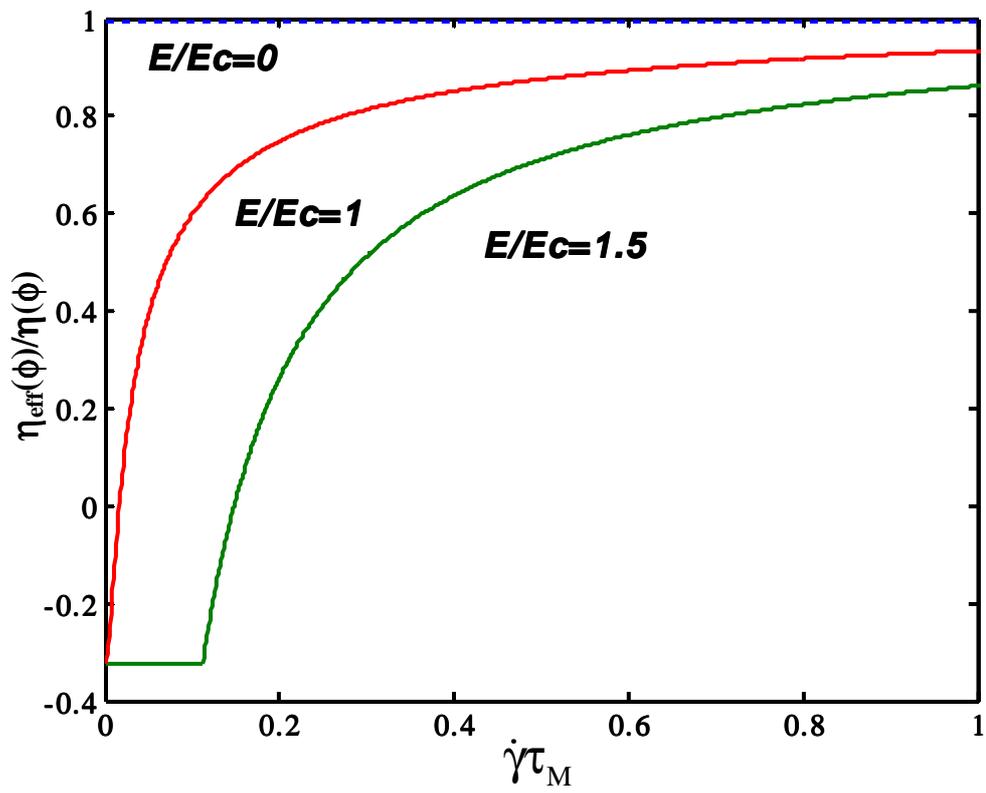

Figure 2a

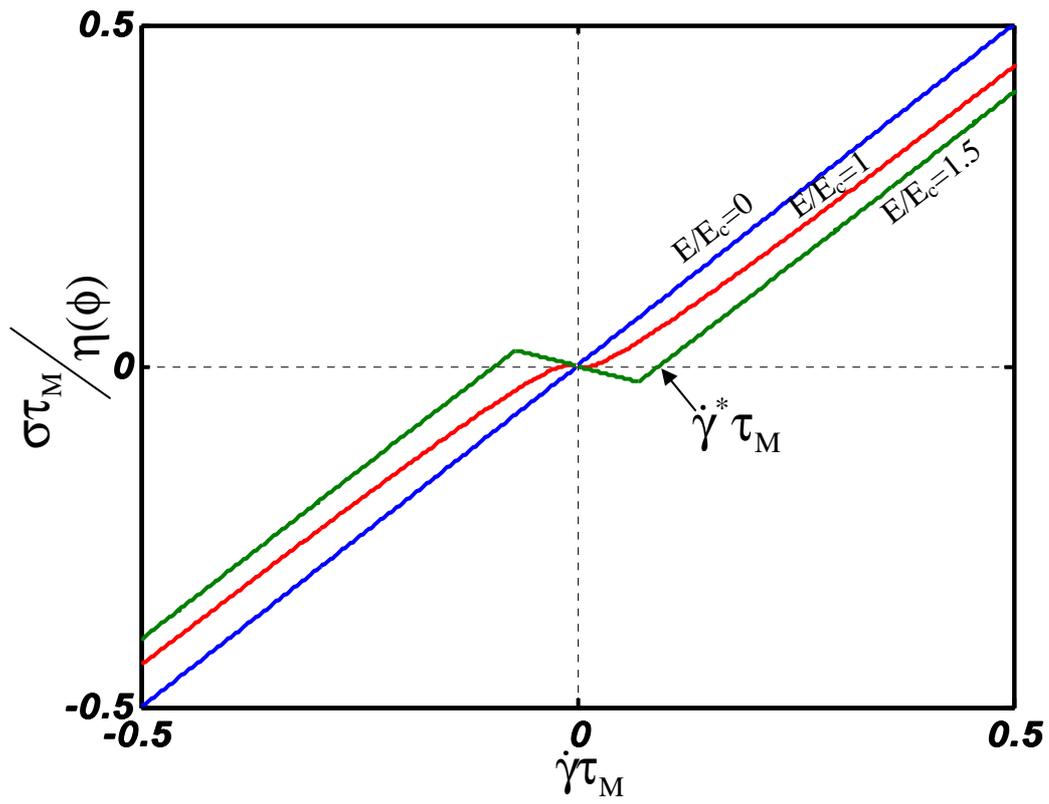

Figure 2.b

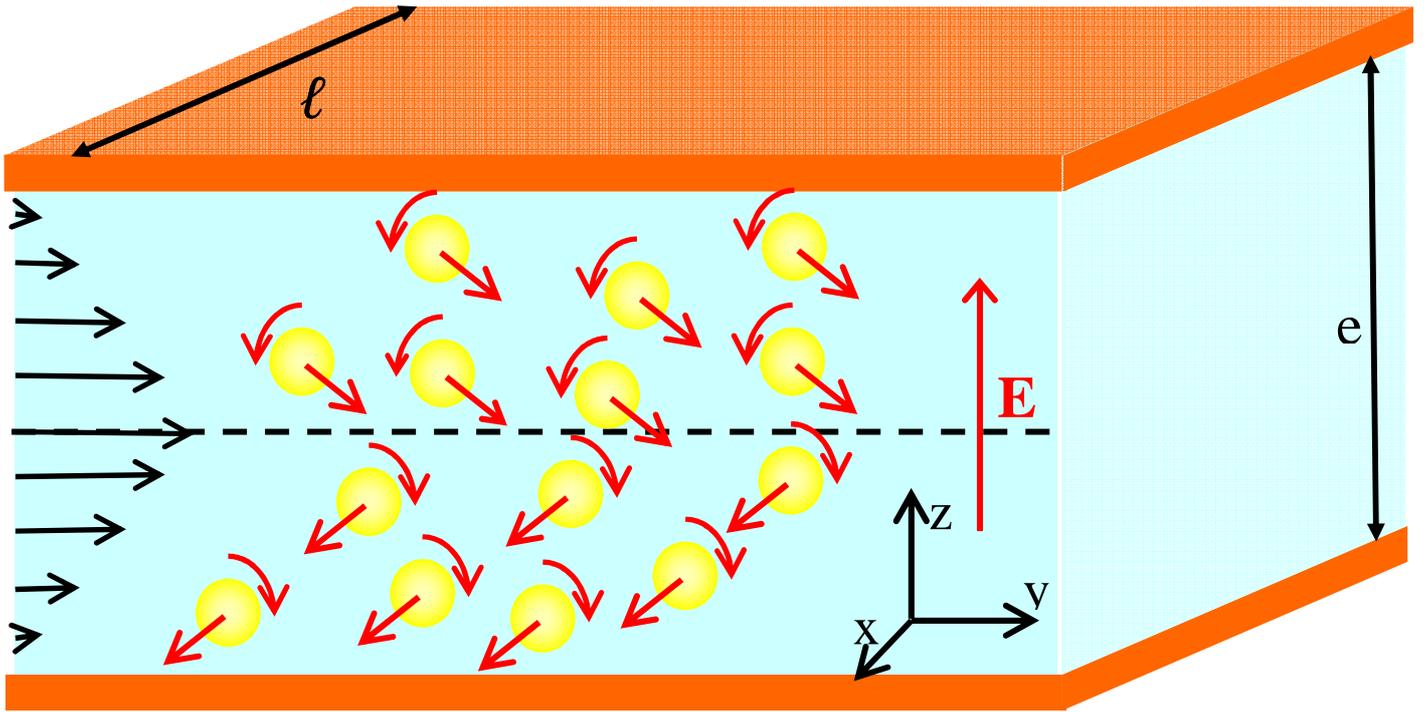

Figure 3

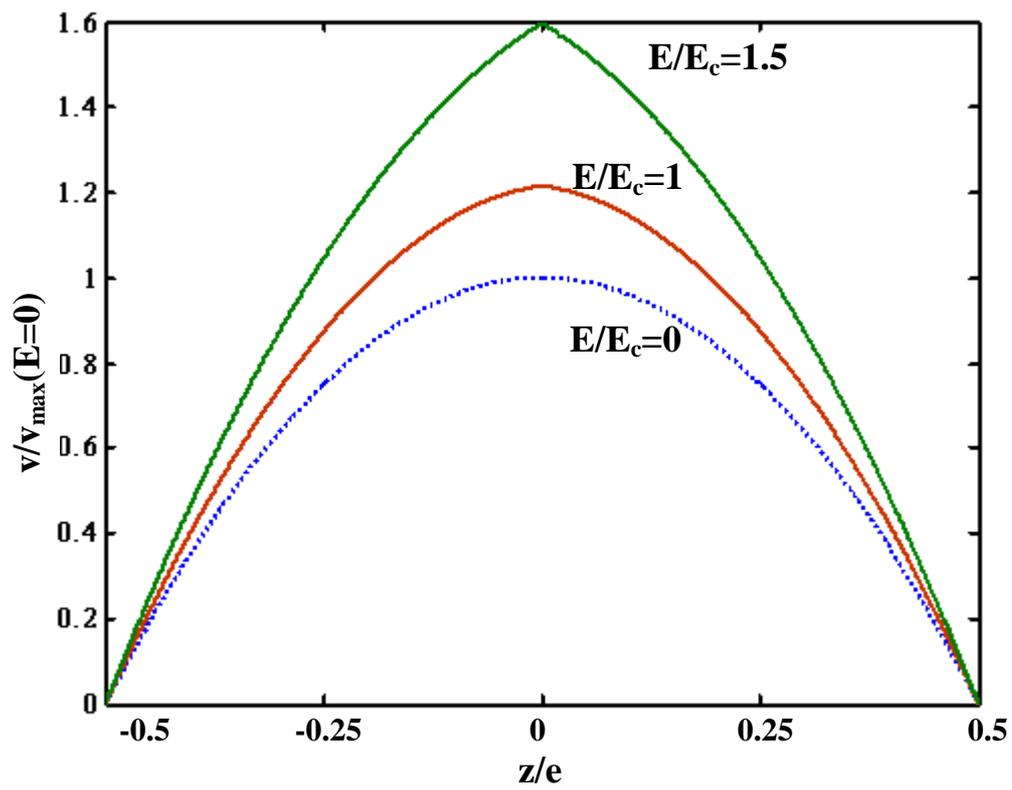

Figure 4

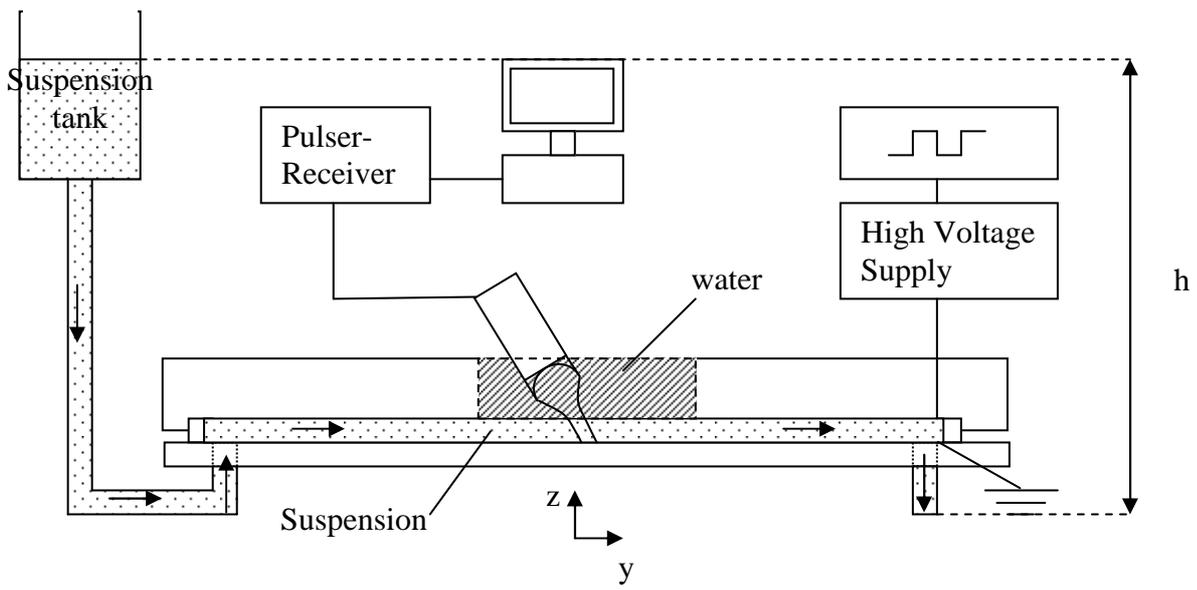

Figure 5

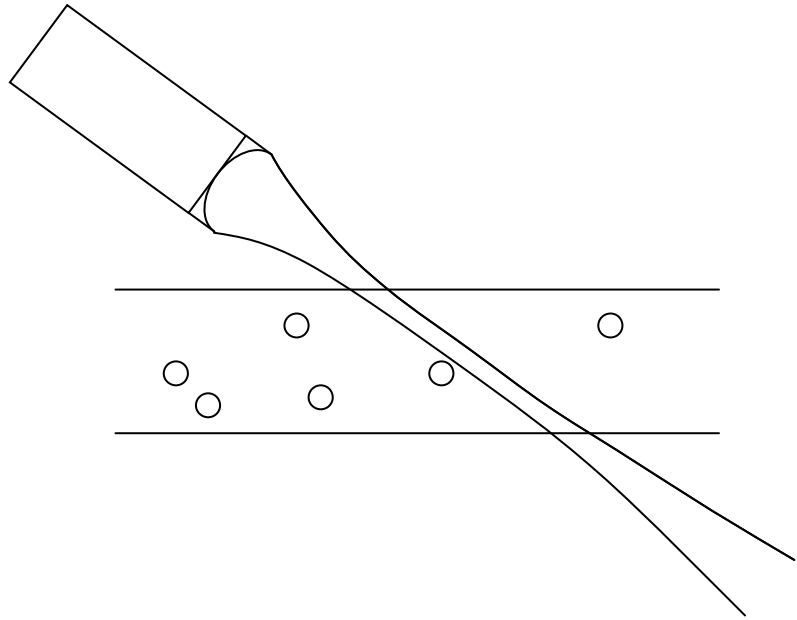

Figure 6.a

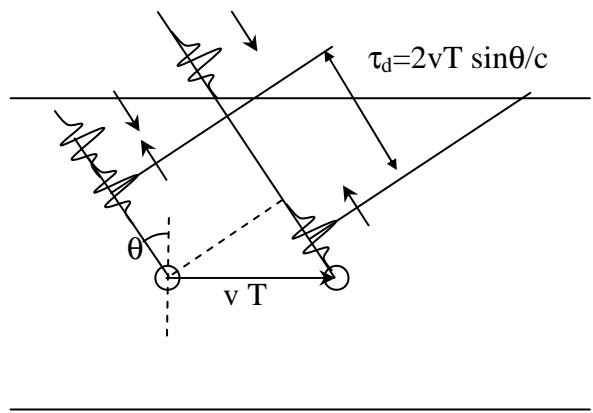

Figure 6.b

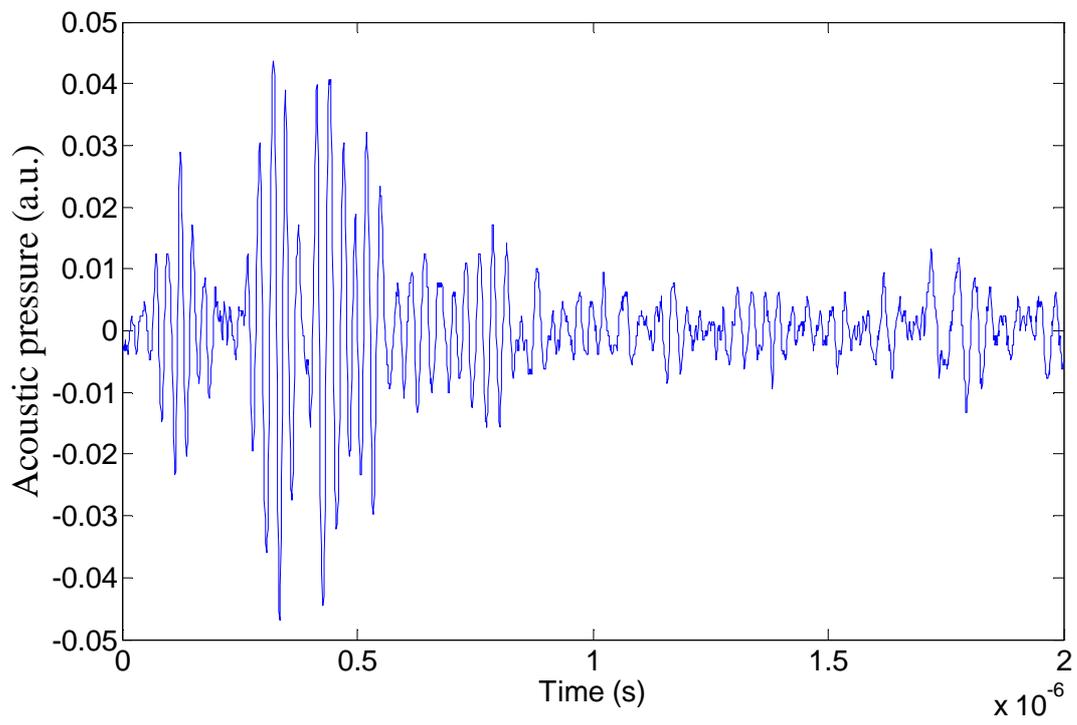

Figure 7

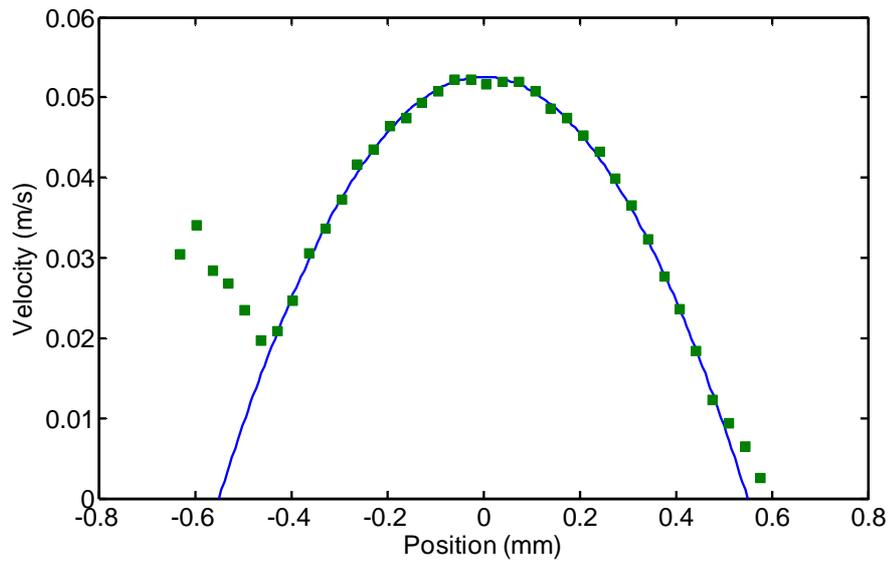

Figure 8

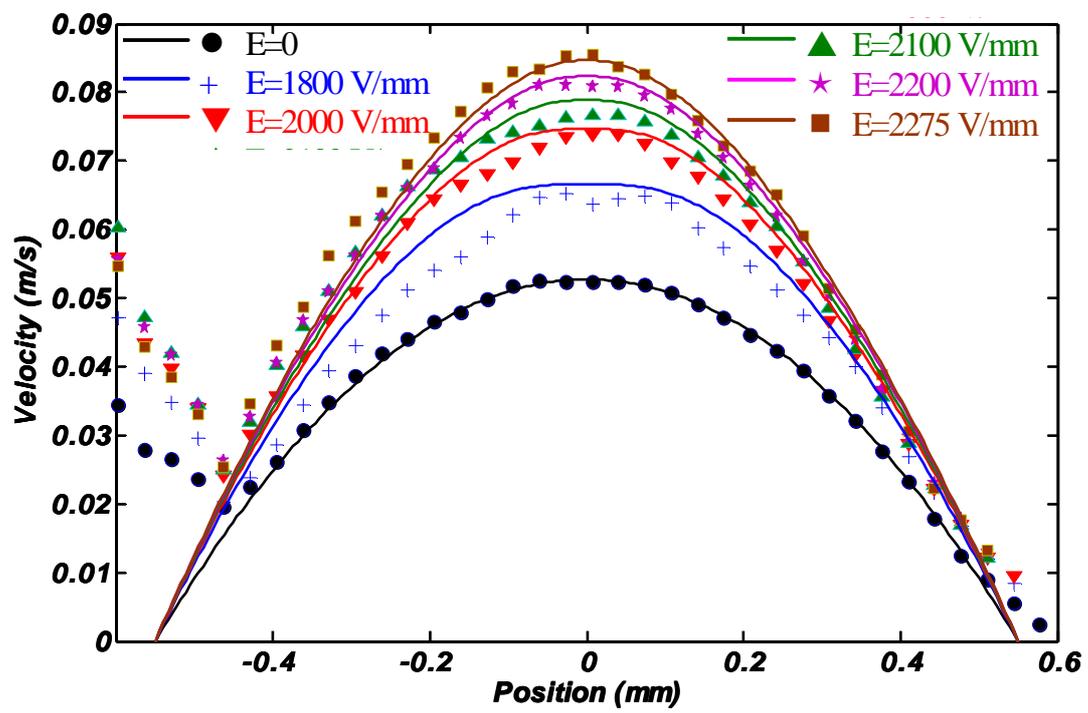

Figure 9

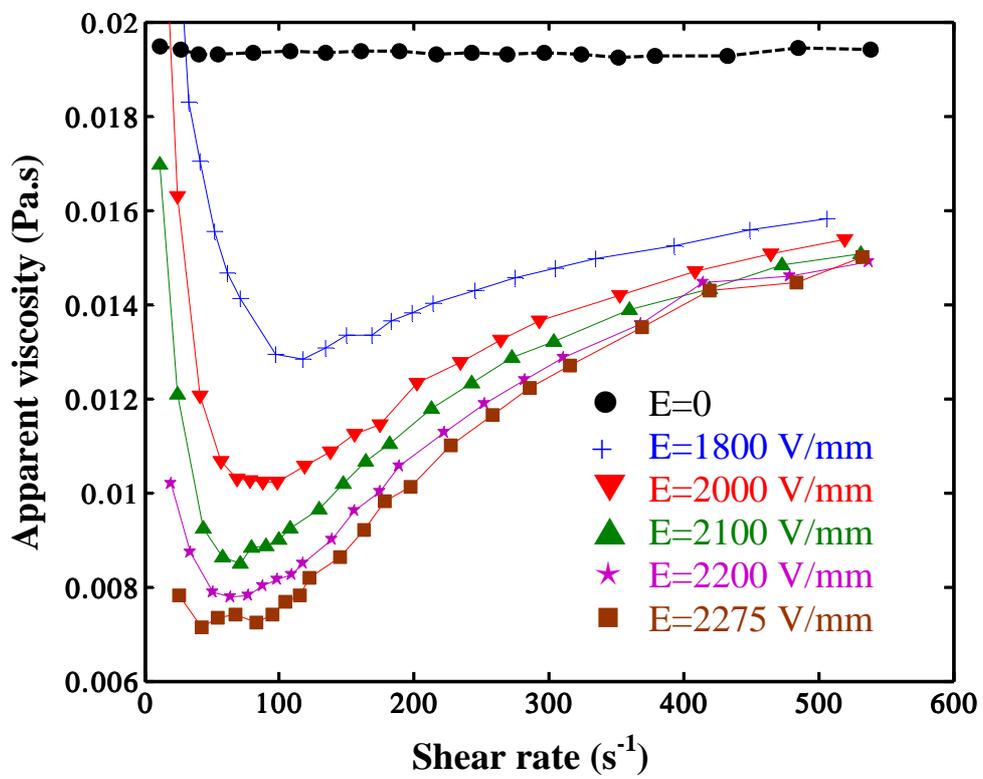

Figure 10